\documentclass[a4paper,showpacs,preprintnumbers,showkeys]{revtex4}

\usepackage{fullpage}
\usepackage{graphicx}% Include figure files
\usepackage{subfigure}
\usepackage{epsfig}
\usepackage{dcolumn}% Align table columns on decimal point
\usepackage{bm}% bold math
\usepackage{amsmath,amssymb,appendix}

\begin{document}

\author{P. S. Pal}
\email{priyo@iopb.res.in}
\address{Institute of Physics, Sachivalaya Marg, Bhubaneswar - 751005, India}

\author{Shubhashis Rana}
\email{shubho@iopb.res.in}
\address{Institute of Physics, Sachivalaya Marg, Bhubaneswar - 751005, India}

\author{Arnab Saha}
\email{sahaarn@gmail.com}
\address{Institut f$\ddot{u}$r Theoretische Physik II, Weiche Materie,
Heinrich-Heine-Universit$\ddot{a}$t D$\ddot{u}$sseldorf,
40225 D$\ddot{u}$sseldorf, Germany.}

\author{A. M. Jayannavar}
\email{jayan@iopb.res.in}
\address{Institute of Physics, Sachivalaya Marg, Bhubaneswar - 751005, India}

\title{Extracting work from a single heat bath - A case study on Brownian particle under external magnetic field in presence of
information }

\author{}

%\date{}

%\maketitle{}
\begin{abstract}
 Work can be extracted from a single bath beyond the limit set by the second law by performing measurement on the system and utilising
the acquired information. As an example we studied a Brownian particle confined in a two dimensional harmonic trap in presence 
of magnetic field, whose position co-ordinates are measured with finite precision. Two separate cases are investigated in this study - (A) moving
the center of the potential and (B) varying the stiffness of the potential. Optimal protocols which extremise the work in a finite time process
are explicitly calculated for these two cases. For Case-A, we show that even though the optimal protocols depend on magnetic field,
surprisingly, extracted work is independent of the field. For Case-B, both the optimal protocol and the extracted work depend 
on the magnetic field. However, the presence of magnetic field always reduces the extraction of work.
\end{abstract}

%\begin{abstract}

%\end{abstract}

\maketitle{}

\newcommand{\nwc}{\newcommand}
\nwc{\vs}{\vspace}
\nwc{\hs}{\hspace}
\nwc{\la}{\langle}
\nwc{\ra}{\rangle}
\nwc{\lw}{\linewidth}
\nwc{\nn}{\nonumber}
\nwc{\tb}{\textbf}
\nwc{\td}{\tilde}

\nwc{\pd}[2]{\frac{\partial #1}{\partial #2}}
\nwc{\zprl}[3]{Phys. Rev. Lett. ~{\bf #1},~#2~(#3)}
\nwc{\zpre}[3]{Phys. Rev. E ~{\bf #1},~#2~(#3)}
\nwc{\zpra}[3]{Phys. Rev. A ~{\bf #1},~#2~(#3)}
\nwc{\zjsm}[3]{J. Stat. Mech. ~{\bf #1},~#2~(#3)}
\nwc{\zepjb}[3]{Eur. Phys. J. B ~{\bf #1},~#2~(#3)}
\nwc{\zrmp}[3]{Rev. Mod. Phys. ~{\bf #1},~#2~(#3)}
\nwc{\zepl}[3]{Europhys. Lett. ~{\bf #1},~#2~(#3)}
\nwc{\zjsp}[3]{J. Stat. Phys. ~{\bf #1},~#2~(#3)}
\nwc{\zptps}[3]{Prog. Theor. Phys. Suppl. ~{\bf #1},~#2~(#3)}
\nwc{\zpt}[3]{Physics Today ~{\bf #1},~#2~(#3)}
\nwc{\zap}[3]{Adv. Phys. ~{\bf #1},~#2~(#3)}
\nwc{\zjpcm}[3]{J. Phys. Condens. Matter ~{\bf #1},~#2~(#3)}
\nwc{\zjpa}[3]{J. Phys. A: Math theor  ~{\bf #1},~#2~(#3)}

\section{Introduction }

 In early nineteenth century, it was realized that conversion of heat into useful work requires two heat baths - a warm source and a cold
sink, between which an engine operates in cycles to convert a portion of heat into work. Laws of macroscopic thermodynamics 
determine the maximum efficiency of the engine under quasistatic conditions \cite{callen}. This efficiency is given by $\eta_C=1-\frac{T_l}{T_h} $.
It follows that no work can be extracted from a single bath when $T_l=T_h$. However, this result 
can be significantly altered if one uses information about the microscopic state of the system as a feedback to its operational mode 
\cite{cao,sagawa10,horo,espo,abreu12,lahiri12,sagawa12,shubho12,shubho13}.
Harnessing information to do useful task is vital in many other disciplines: in biology cellular organisms use the information about 
their environment as a feedback to adapt themselves, in engineering sciences often input of a dynamical system is manipulated by 
feeding back the information from its output for greater stability \cite{holland1,holland2}. In physics, Szilard engine \cite{leo} demonstrates how possession of 
information can lead to extraction of work from a single heat bath without violating second law. Recent techniques of handling 
single molecule provide us the scope to explore such systems in practice. In fact, experimentally it is demonstrated that information
can be converted into useful energy using a colloidal particle trapped by two feedback controlled electric fields \cite{toyabe}. 

%In this case work can be extracted from a single bath beyond the limit set by second law.
In the presence of information second law is modified as
\begin{equation}
 W\geq\Delta F-I,
\label{2nd}
\end{equation}
where $I$ is the mutual information. $I$ being a positive quantity, work performed on the thermodynamic system can be lowered by feedback control \cite{abreu11,bauer}. 
Work can also be extracted from a single bath when the system is driven out of equilibrium. This is the case for molecular motors/ratchets in 
presence of load \cite{reimann,mcm}. 

In this article we explore how to extract work, utilising information, from 
a system driven out of equilibrium but always being attached to a single bath using optimal protocol in presence of static magnetic field. Our system 
consists of a single Brownian particle in a two dimensional harmonic trap. In section \ref{dyna} we describe our system and its dynamics. In sections 
\ref{mov} and \ref{tds} we obtain analytical results for Case-A and Case-B, mentioned in the abstract. Finally we conclude in section \ref{con}. Each section is made self-contained. 
\section{System and its Dynamics }
\label{dyna}

We consider a system consisting of a charged Brownian particle of mass $m$ and charge $q$, constrained to move on a two dimensional (X-Y) plane
under the influence of a time dependent two dimensional harmonic potential and a constant magnetic field $\vec{B}=B\hat{k}$ perpendicular to that plane. We consider two
 different protocols to drive the system out of equilibrium - Case-A. moving the minima of the harmonic trap with constant stiffness and Case-B. changing 
the stiffness of the harmonic trap with time keeping the minima fixed. In both the cases, we first measure the position of the particle and then using 
the information gained from the measurement, we apply time dependent protocols to extract work out of it. The initial measurement accompanied by feedback 
 is responsible for extracting work from the system even if it is attached to a single heat bath for all the time, %By repeated 
%measurement accompanied by feedback we can have information machine from a single bath,
 thereby converting information into work. In this article our aim is to show the influence of the constant magnetic field on 
the optimally extracted work from the system.

Work done on similar systems had been calculated in the overdamped as well as the underdamped regimes in \cite{mamta,arnab}. Work distributions had been 
studied for various protocols. Using Jarzynski equality it had been shown that though the distribution of work depends on
the magnetic field, the free energy is not\cite{mamta,arnab,jayan07,acquino08}. One can write the model Hamiltonian of such systems when isolated from the bath as
\begin{equation}
 H=\frac{1}{2m}\left[\left(p_x+\frac{qBy}{2}\right)^2+\left(p_y-\frac{qBx}{2}\right)^2\right]+\frac{1}{2}k(t')[(x-\alpha_x(t'))^2+(y-\alpha_y(t'))^2]
\end{equation}
where $k(t')$ and $(\alpha_x(t'),\alpha_y(t'))$ are the stiffness and minima of the harmonic trap respectively. Being time dependent they act as protocol 
to drive the system out of equilibrium. We have chosen a symmetric gauge producing a constant magnetic field $B$ along $z$-direction.
The influence of the Lorentz force and the time dependent harmonic trap on the Brownian particle is modeled by the following 
underdamped Langevin equation as 
\begin{equation}
 m\ddot{x}=-\gamma \dot{x}+\frac{qB}{c}\dot{y}-k(t')\{x-\alpha_x(t')\}+\eta_x(t')
\label{un-lan1}
\end{equation}
\begin{equation}
 m\ddot{y}=-\gamma \dot{y}-\frac{qB}{c}\dot{x}-k(t')\{y-\alpha_y(t')\}+\eta_y(t').
\label{un-lan2}
\end{equation}
Double and single dots over $x$ and $y$ imply double and single derivative with respect to $t'$.
Here $\eta_x$ and $\eta_y$ are the components of the thermal noise from the bath in $x$ and $y$ directions. 
The mean value of the Gaussian noise is zero and they are delta correlated with $\langle \eta_i(t')\eta_j(t'')\rangle=2D\delta_{ij}\delta(t'-t'')$ 
for $i,j=x,y$. The strength of the noise $D$, friction coefficient $\gamma$ and temperature $T$ of the bath are related to each other by
 the usual fluctuation dissipation relation, i.e., $D=\gamma k_B T$, where $k_B$ is the Boltzmann constant.
\section{Case-A: moving trap with constant stiffness }
\label{mov}

In this case, we apply the protocol by shifting the center of the trap while the stiffness is kept fixed at $k(t')=k$.
We restrict our study to the overdamped limit of Eq.(\ref{un-lan1}) and Eq.(\ref{un-lan2}) 
\begin{equation}
 \dot{x}-C\dot{y}=-\frac{k}{\gamma}\{x-\alpha_x(t')\}+\frac{\eta_x(t')}{\gamma},
\label{ov-lan1}
\end{equation}
\begin{equation}
 \dot{y}+C\dot{x}=-\frac{k}{\gamma}\{y-\alpha_y(t')\}+\frac{\eta_y(t')}{\gamma}.
\label{ov-lan2}
\end{equation}
where $C=\frac{qB}{\gamma c}$ is a dimensionless parameter. The Smoluchowski equation associated with the above stochastic dynamics is
\begin{eqnarray}
\frac{\partial P}{\partial t'}=\vec{\nabla}.[\Lambda(\vec{r}-\vec{\alpha}) P +\lambda \vec{\nabla} P] =- \vec{\nabla} . \vec{J},
\label{fok}
\end{eqnarray}
where $P(x,y,t')$ is the probability distribution function(PDF) for the position of the particle, $ {\lambda}=D/\gamma^2(1+C^2)$
 and $\vec{J}$ denotes the current density. $\Lambda$ is a $2\times2$ matrix given by 
 \[{ \Lambda}= \frac{k}{\gamma_e}
\left(\begin{array}{cc}
  1 & C  \\ 
  -C  & 1
   \end{array}
\right)  
\]
with $\gamma_e=\gamma(1+C^2) $. The exact solution of the above equation is obtained in {\cite{acquino09}}.
If initially the system is prepared in an athermal condition it can be shown that as time evolves, the system approaches to an equilibrium 
state and the corresponding distribution is given by
\begin{equation}
P(\vec{r})=\frac{k}{2 \pi k_B T} \exp\left[-\frac{k}{2 k_B T}(x^2+y^2)\right].  
\label{edist}
 \end{equation}
Note that, $P(\vec{r})$ is independent of the magnetic field which is consistent with the Bohr-van Leeuwen theorem 
\cite{jayan81,bohr} on the absence of diamagnetism in classical systems, i.e., free energy evaluated from Eq.(\ref{edist}) is
 independent of the magnetic field. Thus system exhibits neither magnetic moment nor magnetic susceptibility. Throughout our 
calculations we consider Eq.(\ref{edist}) as our initial distribution.
\subsection*{Measurement }

At $t'=0$, we instantaneously measure the position of the particle and it is found to be at ($x_m,y_m$) while it's actual 
position is $(x,y)$. The distribution of classical error in the measurement process is considered to be uncorrelated Gaussian with width $\sigma$.
Hence the conditional probability of ($x_m,y_m$) given ($x,y$) is

\begin{equation}
P(\vec{r}_m|\vec{r})=\frac{1}{2\pi\sigma^2}\exp\left[ -\frac{1}{2\sigma^2}\left\{(x_m-x)^2+(y_m-y)^2\right\}\right],
\end{equation}
where $\vec{r}_m$ and $\vec{r}$ denote $(x_m,y_m)$
and $(x,y)$ respectively. The probability distribution $P(\vec{r})$ just before the measurement is given by Eq.(\ref{edist}). The probability density of measurement
outcome  $(x_m,y_m)$ is 
\begin{eqnarray}
 P(\vec{r}_m)&=&\int  P(\vec{r}_m|\vec{r})P(\vec{r})d^2\vec{r}\nonumber\\
&=&\frac{1}{2\pi \sigma_1^2}\exp\left[-\frac{1}{2\sigma_1^2 }(x_m ^2 + y_m ^2)\right],
\label{measurement}
\end{eqnarray}
where $\sigma_1^2\equiv\sigma^2+\frac{k_BT}{k}$.
Using Bayes' theorem, $P(\vec{r}|\vec{r}_m)P(\vec{r}_m)=P(\vec{r}_m|\vec{r})P(\vec{r})$, we obtain the conditional PDF for true position $(x,y)$
given $(x_m,y_m)$  as
\begin{eqnarray}
 P(\vec{r}|\vec{r}_m)=\frac{1}{2 \pi \beta^{-1}(0)}\exp\left[-\frac{1}{2\beta^{-1}(0)}\left\{(x-A_x(0))^2 +(y-A_y(0))^2\right\}\right],\hspace{1 cm}
\label{ini-dist}
\end{eqnarray}
where $\beta^{-1}(0)=\frac{k_B T \sigma^2}{k\sigma_1^2}=l\sigma^2$ is the initial width after measurement.
$A_x(0)=\frac{x_m k_B T}{k\sigma_1^2}=lx_m$ and $A_y(0)=\frac{y_m k_B T }{k\sigma_1^2}=ly_m$ are the initial
mean along $x$ and $y$ direction respectively with $l\equiv\frac{k_BT}{k\sigma_1^2}$. The distribution in Eq.(\ref{ini-dist}) is the {\it effective} 
initial distribution after measurement. Note that, due to measurement the width of the effective distribution becomes lesser compared to
both the thermal width and the error width.

 A quantity of particular interest in our problem is 
the Kullback-Leibler(K-L) distance or the relative entropy between  the distributions $P(\vec{r}|\vec{r}_m)$ and $P(\vec{r})$
\begin{eqnarray}
 I(\vec{r}_m)&&=\int P(\vec{r}|\vec{r}_m)\ln\left[ \frac{P(\vec{r}|\vec{r}_m)}{P(\vec{r})}\right]d^2\vec{r}\nonumber\\
&&=-\ln\left[\frac{\beta^{-1}(0)}{k_B T/k}\right] + \frac{k}{2k_B T}[A_x ^2(0)+A_y ^2 (0)] +\frac{k\beta^{-1}(0)}{k_B T}-1.
\label{inform}
\end{eqnarray}
 This distance quantifies the distinguishability between two distributions
for a particular measured outcome $(x_m,y_m)$. It contains information gained after measurement. The average of K-L distance over all measured outcomes 
gives the mutual information. Now the mutual information $I$ is related to $ I(\vec{r}_m,\vec{r})=\ln \left[\frac{P(\vec{r}_m,\vec{r})}{P(\vec{r}_m)P(\vec{r})}\right]$ as 
\cite{cover}
\begin{eqnarray}
I&=&\int P(\vec{r}_m)I(\vec{r}_m)d^2\vec{r}_m\nn\\
&=&\int P(\vec{r}_m)\left(\int P(\vec{r}|\vec{r}_m)\ln \left[\frac{P(\vec{r}|\vec{r}_m)}{P(\vec{r})}\right]d^2\vec{r}\right)d^2\vec{r}_m \nn\\ 
&=&\int P(\vec{r}_m,\vec{r}) \ln \left[\frac{P(\vec{r}_m,\vec{r})}{P(\vec{r}_m)P(\vec{r})}\right]d^2\vec{r}_m d^2\vec{r}\nn\\
&=&\langle I(\vec{r}_m,\vec{r})\rangle,\nn
\end{eqnarray}
where we have used Bayes' theorem in the third step. In our case $I$ is given by  
\begin{equation}
 I=\ln\left[1+\frac{k_B T}{k\sigma^2}\right].
\label{inform1}
\end{equation}
%

%%%%%%%%%%%%%%%%%%%%%%%%%%%%%%%%%5
\section*{Instantaneous Process }
%%%%%%%%%%%%%%%%%%%%%%%%%%%%%%

In this section we calculate the work done on the particle for an instantaneous shift of the potential.
We first measure the position of the particle and apply feedback by shifting the  potential minima  from (0,0) to 
$(\alpha_{xf}(x_m),\alpha_{yf}(y_m))$ according to measurement outcome ($x_m,y_m$) instantaneously. For this 
process the work done on the system is the change in internal energy of the system
\begin{equation}
w(\vec{r}_m;\vec{r})=\frac{k}{2}[(x-\alpha_{xf}(x_m))^2-x^2]+\frac{k}{2}[(y-\alpha_{yf}(y_m))^2-y^2].\nonumber
\end{equation}
Averaging over all $\vec{r}$  for fixed $\vec{r}_m$ we have
\begin{eqnarray}
 W(\vec{r}_m)
&=&\int P(\vec{r}|\vec{r}_m)w(\vec{r}_m;\vec{r}) d^2\vec{r}\nonumber\\
&=&\frac{k}{2}\left[(A_x(0)-\alpha_{xf})^2 + (A_y(0)-\alpha_{yf})^2\right]-\frac{k}{2}\left[A^2_x(0)+A^2_y(0)\right],
\end{eqnarray}
which is minimum at $  \alpha_{xf}^*(x_m)=A_x(0)$ and $\alpha_{yf}^*(y_m)=A_y(0)$ and the minimum value of extracted work is given by
 \begin{equation}
 W^*(\vec{r}_m)=-\frac{k}{2}\left[A^2_x(0)+A^2_y(0)\right].
\end{equation}
 Using the expression of information 
$I(\vec{r}_m,\vec{r})$, we can write
\begin{eqnarray}
 \langle \exp\left[-\beta w(\vec{r}_m,\vec{r})-I(\vec{r}_m,\vec{r})\right] \rangle
&=&\int \exp\left[-\beta w(\vec{r}_m,\vec{r})-I(\vec{r}_m,\vec{r})\right] P(\vec{r}_m,\vec{r})d^2\vec{r}_m d^2\vec{r}\nn\\
&=& \int e^{-\beta w(\vec{r}_m,\vec{r})}P(\vec{r}_m)P(\vec{r})d^2\vec{r}_m d^2\vec{r}.\nn
\end{eqnarray}
The expression for $w(\vec{r}_m,\vec{r})$ is substituted in the above equation and it, being a Gaussian integral, can be easily
integrated thereby leading to the desired result
\begin{eqnarray}
 \langle \exp{\left[-\beta w(\vec{r}_m,\vec{r})-I(\vec{r}_m,\vec{r})\right]} \rangle=1.
\end{eqnarray}
This is the modified Jarzynski Equality \cite{sagawa10,shubho12} in presence of information and feedback for an instantaneous shift
 of the potential. Applying Jensen's inequality we obtain
\begin{equation}
\beta \langle w(\vec{r}_m,\vec{r})\rangle \ge -\langle I(\vec{r}_m,\vec{r})\rangle=-I,
\end{equation}
which is the modified Second Law in presence of information. Thus the maximum work extracted from the system is bounded by the information we obtain
by measurement.

%%%%%%%%%%%%%%%%%%%%%%%%%%%%%%%
\subsection*{Calculation of optimal work for optimal protocols in finite time process}
%%%%%%%%%%%%%%%%%%%%%%%%%%%%%%%%%%%%

In this process we assume the system to be initially in equilibrium at $t'=0$ and a measurement is done at that time. Depending on the  measurement outcomes, a 
protocol is applied as a feedback to the system for a finite time. The protocol shifts the center of the trap from initial position $(0,0)$
to final position $(f_1,f_2)$ in a total time $t$. We calculate the work done on the particle during this process using the definition
 of the thermodynamic work as given by Jarzynski \cite{jarzynski}. Thus
\begin{equation}
 w=\int_0^t \frac{\partial V}{\partial \alpha_x} \dot{\alpha_x}dt'+\int_0^t \frac{\partial V}{\partial \alpha_y} \dot{\alpha_y} dt',
\end{equation}
where $V=\frac{1}{2}k[(x-\alpha_x(t'))^2+(y-\alpha_y(t'))^2]$ is the confining potential. Averaging over all possible 
realizations of Gaussian noise, the work becomes 
\begin{equation}
 W(\vec{r}_m)=\la w \ra=\int_0^t\left\la \frac{\partial V}{\partial \alpha_x}\right\ra \dot{\alpha_x}dt'+\int_0^t\left\la \frac{\partial V}{\partial \alpha_y}\right\ra \dot{\alpha_y} dt'.\nonumber\\
\end{equation}
After some straight forward algebra we get
\begin{eqnarray}
W(\vec{r}_m)&=&\frac{1}{2}kf_1^2-kA_x(t)f_1+k\int_0^t\dot{A_x}(t')\alpha_x(t')dt'+\frac{1}{2}kf_2^2\nonumber\\
&&\hspace{3 cm}-kA_y(t)f_2+k\int_0^t\dot{A_y}(t')\alpha_y(t')dt',
\label{avg-W1} 
\end{eqnarray}
where $A_x(t')=\la x(t')\ra$ and $A_y(t')=\la y(t')\ra$. Taking noise average on both sides of 
Eq.(\ref{ov-lan1}) and Eq.(\ref{ov-lan2}) and after rearranging one can write
\begin{equation}
 \alpha_x(t')=\frac{\gamma}{k}\{\dot{A_x}(t')-C\dot{A_y}(t')\}+A_x(t')
\label{alpha_x}
\end{equation}
\begin{equation}
\alpha_y(t')=\frac{\gamma}{k}\{C\dot{A_x}(t')+\dot{A_y}(t')\}+A_y(t').
\label{alpha_y}
\end{equation}
Replacing the above expressions for $\alpha_x(t')$ and $\alpha_y(t')$ in Eq.(\ref{avg-W1}), the average work can be expressed as a sum
of a boundary term and an integral term:
\begin{equation}
  W(\vec{r}_m)=\frac{1}{2}k[(A_x(t)-f_1)^2+(A_y(t)-f_2)^2]-\frac{1}{2}k[A_{x}^2(0)+A_{y}^2(0)]+\gamma\int_0^t(\dot{A_x}^2+\dot{A_y}^2)dt^\prime.
\label{f1-w}
\end{equation}
We now evaluate the optimal protocol that extremises the work. Here we follow the same procedure adopted in \cite{abreu11}.
Note that, the work can be expressed as $W=W_x+W_y$, where
\begin{equation}
 W_x=\frac{1}{2}k\{A_x(t)-f_1\}^2-\frac{1}{2}kA_{x}^2(0)+\gamma\int_0^t\dot{A_x}^2dt^\prime,
\label{wx}
\end{equation}
\begin{equation}
 W_y=\frac{1}{2}k\{A_y(t)-f_2\}^2-\frac{1}{2}kA_{y}^2(0)+\gamma\int_0^t\dot{A_y}^2dt^\prime.
\label{wy}
\end{equation}
In case of $W_x$, extremising the integral part by variational principle we obtain the Euler-Lagrange equation for $A_x(t')$. Solving
this equation we find that $A_x(t')$ is linear in time and is given by 
\begin{equation}
A_x(t')=c_1t'+A_x(0),
\label{a_x}
\end{equation}
 with, $c_1=\frac{A_x(t)-A_x(0)}{t}$.
When $W_x$ is further extremised with respect to the final value, $A_x(t)$, we get  
\begin{equation}
 A^*_x(t)=\frac{kf_1t+2\gamma A_x(0)}{kt+2\gamma}.
\end{equation}
Using Eq.(\ref{a_x}) we get 
\begin{equation}
A^*_x(t')=\left[\frac{A_x^*(t)-A_x(0)}{t}\right]t'+A_x(0).
\end{equation}
Similarly for extremisation of $W_y$, the corresponding equation is given by 
\begin{equation} 
A^*_y(t')=\left[\frac{A_y^*(t)-A_y(0)}{t}\right]t'+A_y(0),
\end{equation}
with 
\begin{equation}
 A^*_y(t)=\frac{kf_2t+2\gamma A_y(0)}{kt+2\gamma}.
\end{equation}
The optimal protocols are obtained by using the above expressions of mean values in Eq.(\ref{alpha_x}) and Eq.(\ref{alpha_y}).
\begin{figure}[h]
 \begin{centering}
  \includegraphics[height=4.5in,width=4.5in]{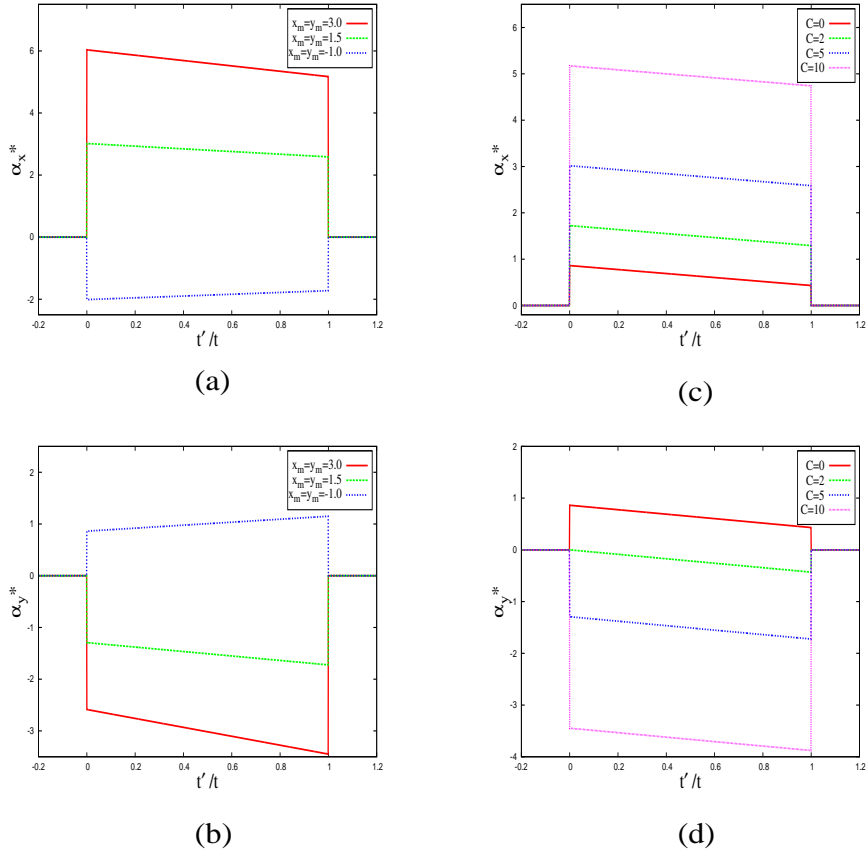}
  \caption{Optimal protocols corresponding to moving trap are plotted for cyclic process, .i.e., $f_1=f_2=0$ in presence of 
measurement for total time $t=1$. The error width is taken to be $\sigma=0.4$. In (a) and (b), the optimal protocols $\alpha_x^*$ and $\alpha_y^*$ are plotted 
for fixed magnetic field $C=5$ but for three different measurement outcomes. In (c) and (d) the same are plotted with a fixed 
measurement outcomes $x_m=y_m=1.5$ but with four different magnetic fields.}
\label{mt-pro}
 \end{centering}
\end{figure}
Replacing the expressions of $A_x^*(t)$ and $A_y^*(t)$ in Eq.(\ref{f1-w}),  we  get the final result for optimal work:
\begin{eqnarray}
 W^*(\vec{r}_m)
=\frac{\gamma k}{(kt+2\gamma)}\{(f_1-A_x(0))^2+(f_2-A_y(0))^2\}-\frac{1}{2}k[A_x^2(0)+A_y^2(0)].
\end{eqnarray}
To obtain average optimal work, $W^*(\vec{r}_m)$ is averaged over all measurement outcomes $\vec{r}_m$ using Eq.(\ref{measurement}) and its 
expression is given by
\begin{eqnarray}
 \overline{W^*}
&=&\int_{-\infty}^{\infty}\int_{-\infty}^{\infty}W^*(\vec{r}_m)P(\vec{r}_m)d^2\vec{r}_m\nn\\
&=&\frac{\gamma k}{(kt+2\gamma)}(f_1^2+f_2^2)-\frac{(k_B T)^2}{\sigma_1^2(kt+2\gamma)}t.
\label{eq:w-1}
\end{eqnarray}
We emphasize that even though optimal protocol depends on the magnetic field, the average work done on the particle
is independent of the magnetic field. This is rather a surprising result. Magnetic field itself does not do any work as 
the Lorentz force is perpendicular to the displacement of the particle. However, magnetic field continually changes the
 direction of the particle thereby changing the work done on the particle by other work sources. For example, the work
source, in our present case, is the protocol that changes the minima of the potential.
From Eq.(\ref{eq:w-1}) we observe that the first part is strictly positive while the second one is strictly negative.
In the $t\rightarrow 0$ limit
\begin{equation}
 \overline {W^*}(t\rightarrow 0)=\frac{k}{2}(f_1^2+f_2^2), 
\end{equation}
which implies we cannot extract any work in this limit. Intuitively it can be understood that for $t\rightarrow0$ case there is only one instantaneous 
jump in the optimal protocol, from $0$ to a fixed point, and hence there is no time to take the advantage from the acquired information.
For large time 
\begin{equation}
 \frac{\overline{ W^*}(t\rightarrow \infty)}{k_BT}= -\frac{k_B T}{k\sigma_1^2}=-\frac{X}{1+X},
\end{equation}
where $X=\frac{k_B T}{k\sigma^2}$.
This quantity being negative, we can always extract work. The magnitude of this work does not depend on the protocol 
parameters. It shows that more inefficient our measurement is, the lesser will be the amount of work extracted. 
\begin{figure}[h]
 \begin{centering}
 \includegraphics[height=2in,width=4.5in]{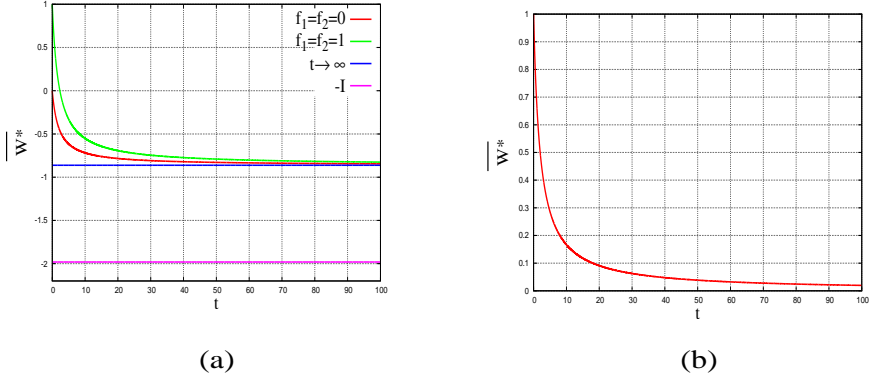}
  \caption{Average extremal work for moving trap (a) with measurement (error width $\sigma=0.4$) (b) without measurement. In (a) we 
plot the extremal work for two different final conditions, its asymptotic value for $t\rightarrow\infty$ limit and the acquired
information $I$ due to measurement.}
\label{avg-w} 
\end{centering}
\end{figure}
The most inefficient measurement ($\sigma\rightarrow\infty$) is equivalent to no measurement. As information obtained from 
most inefficient measurement tends to zero [evident from Eq.(\ref{inform1})], we cannot extract work in this limit. We also conclude from Eq.(\ref{eq:w-1}) that, for cyclic 
protocols $(f_1=f_2=0)$, work is always extracted.

In the absence of any measurement, the protocols are given by 
\begin{eqnarray}
 \alpha_x^*(t')&=&\frac{\gamma(f_1-Cf_2)}{kt+2\gamma}+\frac{kf_1}{kt+2\gamma}t',\\
\alpha_y^*(t')&=&\frac{\gamma(Cf_1+f_2)}{kt+2\gamma}+\frac{kf_2}{kt+2\gamma}t'.
\end{eqnarray}
The corresponding optimal work is 
\begin{equation}
 \overline{W^*}=\frac{\gamma k}{(kt+2\gamma)}(f_1^2+f_2^2),
\end{equation}
which is positive and independent of magnetic field. Hence work cannot be extracted. In contrast to the above results
we have verified that for non-optimal protocols, work depends on magnetic field.
In Fig.(\ref{mt-pro}a) and (\ref{mt-pro}b) we have plotted $\alpha_x^*(t')$ and $\alpha_y^*(t')$ for fixed magnetic field and 
different values of measurement position ($x_m$ and $y_m$). In Fig.(\ref{mt-pro}c) and (\ref{mt-pro}d) we have plotted
 the optimal protocols for different values of magnetic field with fixed measurement positions. The optimal protocols
 show discontinuities at the initial and end points. This is the generic feature of optimal protocols \cite{schmiedl,abreu11}.

In Fig.(\ref{avg-w}a) we have plotted the average optimal work as function of time for different 
protocols whose end points are $f_1=f_2=0$ and $f_1=f_2=1$. The extracted work saturates to the limit $-\frac{k_BT}{\sigma_1^2}$ which is independent of $f_1$ and $f_2$.
However, it is less than the information gain $I$(shown on the graph). In Fig.(\ref{avg-w}b) we have plotted average work for optimal 
protocols in the absence of measurements. It is clear from this figure that $\overline{W^*}$ is always positive and one cannot
extract any work in the absence of information and feedback. In the next section we study the optimal work extraction when the stiffness of the 
trap is varied with time.
%%%%%%%%%%%%%%%%%%%%%%%%%%%
\section{Case-B: time dependent stiffness }
\label{tds}
%55555555555555555555555555

\begin{figure}[h]
 \begin{centering}
  \includegraphics[height=4.5in,width=4.5in]{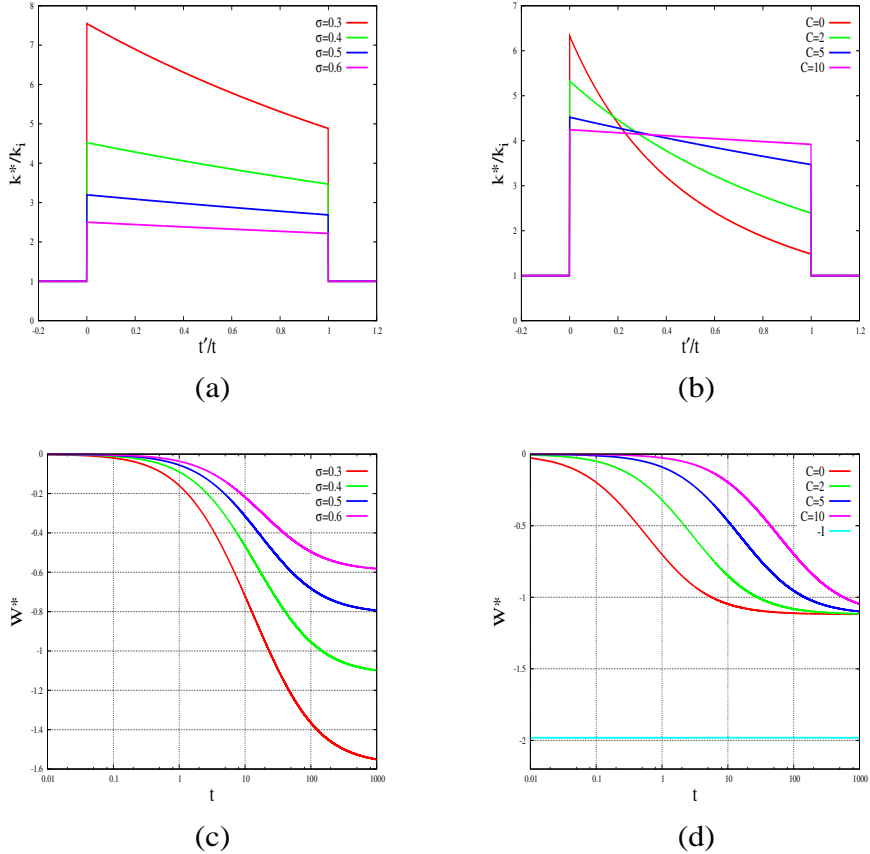}
  \caption{Optimal protocol and work for stiffness varying trap are plotted in presence of measurement for cyclic process in 
total time $t=1$. The process being cyclic, the initial and final stiffness are same and chosen to be unity. In (a) and (c) 
optimal protocol and work are plotted for fixed magnetic field $C=5$ wheras in (b) and (d) the same are plotted for fixed 
error width $\sigma=0.4$. In (c) and (d) log time time scale is used to depict the fact that $W^*$ saturates at large times and
the saturation value is independent of magnetic field.}
  \label{k-me} 
 \end{centering}
\end{figure}

\begin{figure}[h]
 \begin{centering}
  \includegraphics[height=2in,width=2in]{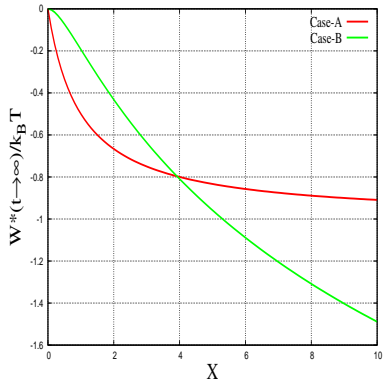}
  \caption{Plot of extracted work in the units of $k_BT$ at $t\rightarrow\infty$ limit for Case-A and Case-B.}
  \label{wstar} 
 \end{centering}
\end{figure}

In this case the stiffness $k(t')$ of the potential is changed from $k(0)=k_i$ to $k(t)=k_f$ 
keeping the center of trap  fixed at $(0,0)$. Initially the particle is in equilibrium with $k=k_i$. Now at $t=0$ a measurement is performed and the distribution just after the measurement 
for a given outcome $(x_m,y_m)$ is given by  Eq.(\ref{ini-dist}). We follow the same procedure for the initial measurement and feedback as in 
Case-A. The overdamped Langevin equations for this case are 
\begin{equation}
 \dot{x}-C\dot{ y}=-\frac{k(t')}{\gamma}{x}+\frac{\eta_x}{\gamma},
\label{xe}
\end{equation}
\begin{equation}
 \dot{ y}+C\dot{x}=-\frac{k(t')}{\gamma}{y}+\frac{\eta_y}{\gamma}.
\label{ye}
\end{equation}
 We rewrite the above equations using the variable $z=x+iy$, ($i=\sqrt{-1}$) and $\eta_z=\eta_x+i\eta_y$ as 
\begin{equation}
\frac{dz}{dt'}+\frac{k(1-iC)}{\gamma_e}z=\frac{1-iC}{\gamma_e}\eta_z,
\end{equation}
 the solution of which is given by
\begin{equation}
z=z_0+\frac{1-iC}{\gamma_e}\exp\left[-\frac{1-iC}{\gamma_e}R(t')\right]\int_0^{t'}\eta_z(t'')\exp\left[\frac{1-iC}{\gamma_e}R(t'')\right]dt'' ,
\label{solz}
\end{equation}
where $R(t')=\int_0^{t'}k(t'')dt''$ and $z_0$ is a constant which is fixed by initial measurement.
Using Eq.({\ref{solz}}) the time evolution of the second moment $\Sigma=\langle x^2\rangle+\langle y^2\rangle=\la zz^*\ra$ can be 
written as  
\begin{equation}
 \dot{\Sigma}=-\frac{2k(t')}{\gamma_e}\Sigma+\frac{4k_BT}{\gamma_e}.
\label{sig}
\end{equation}
The expression of work is given by 
\begin{equation}
w=\frac{1}{2}\int_0^t \dot{k}(x^2+y^2)dt'.
\end{equation}
The average work can be written as a functional of $\Sigma$ and its derivative 
\begin{equation}
 W=\la w\ra=\int_0^{t}dt'\dot k \frac{\Sigma}{2}=\frac{1}{2}[k\Sigma-2k_BT\ln\Sigma]_0^t+\frac{\gamma_e}{4}\int_0^tdt^{\prime}\frac{\dot{\Sigma}^2}{\Sigma}.
\label{work-stiff}
\end{equation}
Here we have first integrated the average work by parts and then substituted $k(t')$ 
from Eq.(\ref{sig}). Extremising the integral part in the expression of work using Euler-Lagrange equation we have 
\begin{equation}
 \Sigma(t^{\prime})=c_1(1+c_2t^{\prime})^2.
\label{sig-sol}
\end{equation}
The initial distribution conditioned to the measurement outcome fixes $c_1=2\beta^{-1}(0)=\frac{2k_BT\sigma^2}{k_i\sigma_1^2}$. 
Substituting the expression for $\Sigma(t')$ in Eq.(\ref{work-stiff}) we get 
\begin{equation}
 W=k_i\beta^{-1}(0)\left[\frac{k_f}{k_i}(1+c_2t)^2-1\right]-2k_BT\ln(1+c_2t)+\frac{2\gamma_e\beta^{-1}(0)}{t}(c_2t)^2.
\label{opt-work}
\end{equation}
Now to obtain the second constant $c_2$ we minimize Eq.(\ref{opt-work}) with respect to $c_2$ leading to an optimal value
\begin{equation}
 c_2^*t=-1+\frac{1+\sqrt{1+\frac{2k_BTt}{\gamma_e \beta^{-1}(0)}(1+\frac{k_ft}{2\gamma_e})}}{2(1+\frac{k_ft}{2\gamma_e})}.
\end{equation}
Substituting the values of $c_1$ and $c_2^*$ in Eq.(\ref{sig-sol}) and making use of Eq.(\ref{sig}) we obtain the optimal protocol 
\begin{equation}
 k^*(t^{\prime})=\frac{k_BT}{\beta^{-1}(0)(1+c_2^*t^{\prime})^2}-\frac{\gamma_e c_2^*}{(1+c_2^*t^{\prime})},
\end{equation}
for $0<t'<t$ and it implies jumps at the beginning and at the end of the process. In Fig.(\ref{k-me}a) we have plotted optimal protocols
as a function of time for fixed magnetic field and different values of measurement error $\sigma$. In Fig.(\ref{k-me}b) we have plotted 
optimal protocols as a function of time for fixed measurement error $\sigma$ and different values of magnetic field. The initial and final jumps 
in the protocol are clearly visible as in Case-A and the protocols are magnetic field dependent.
\\In Fig.(\ref{k-me}c), we have plotted optimal work done on the particle in a cyclic process, .i.e., $k_i=k_f=k$ as a function of time for 
given magnetic field  and different measurement errors while in Fig.(\ref{k-me}d) optimal work is plotted for a fixed measurement error
and different magnetic fields.
From Fig.(\ref{k-me}c) it is clear that, unlike Case-A, the optimal work in finite time process depends on the magnetic field. It decreases with time and saturates
to a value which independent of the magnetic field and is given by
\begin{equation}
 \frac{W^*({t\rightarrow\infty})}{k_BT}=\frac{X}{1+X}-\ln(1+X).
\label{w-sat}
\end{equation}
 It may be noted that for all values of $X$, $\frac{W^*(t\rightarrow\infty)}{k_BT}$ is negative. It can be shown analytically by noting the fact that 
it has negative slope  for all $X$ and a maximum value of zero at $X=0$ as shown in Fig.(\ref{wstar}). 
This is true only for cyclic processes. The approach of $W^*(t)$ towards $W^*(t\rightarrow\infty)$ is controlled by the
 applied magnetic field.  
This is in sharp contrast to the result we obtained for moving trap. 
The plot of optimal work, in presence of magnetic field, always lies above that in absence of magnetic field. 
In Fig.(\ref{k-me}d), we observe that the measurement error decreases the work extraction as expected. While comparing the extracted work at large time for 
both the protocols, discussed earlier, we see from Fig.(\ref{wstar}) that, if the ratio of thermal and error width exceeds a threshold ($X=3.92086$), one can extract more 
work by varying the stiffness of the trap optimally with time than the work extracted by optimally moving the  trap.

%%%%%%%%%%%%%%%%%%%%%%%%%
\section{Conclusions}
\label{con}
%%%%%%%%%%%%%%%%%%%%%%%%%%

In this section we summarize our results. We carried out an analytical study of optimal protocols using variational principle in presence of measurement 
and feedback. Using this, the work extracted by our system in presence of single bath for finite time process is obtained. This depends on the initial measurement
 of the position of the particle and the acquired information. For Case-A optimal protocols depend on magnetic field, measured co-ordinates
of the particle and measurement error. However for Case-B, the optimal protocol is independent of the measured positions. In both cases, 
information helps in extracting work from the system, but for Case-A it is magnetic field independent whereas in Case-B it depends 
on the magnetic field. The saturated value of work extraction is bounded by mutual information. As a special case of moving trap, 
Jarzynski identity in presence of information has been verified for an instantaneous change in the protocols.

Finally, we would like to emphasize the following points. It is to be noted that in our treatment, we perform only one measurement of the 
co-ordinates of the particle at the start of the process $(t'=0)$. This measurement changes the distribution of the particle position
at time $t'=0$ to an athermal distribution as given in Eq.(\ref{ini-dist}). It is known that, from athermal distribution one can 
always extract work - bounded by the information measure $I$ which is the K-L distance between athermal initial distribution and
 the corresponding equilibrium distribution \cite{espo,vaikunta,lahiri14}. From the above discussion it is evident that performing a single measurement 
starting from a thermally equilibrated state is equivalent to starting with a nonequilibrium distribution. Further investigations are being carried out 
in this direction.

\section{ACKNOWLEDGMENTS}
A.M.J. thanks DST, India for financial support and. A.S. thanks MPIPKS, Germany for partial support.

\end{document}